\documentstyle[pra,aps,twocolumn]{revtex}
\begin{document}
\draft
\title{On the volume of the set of mixed entangled states}

\author{Karol \.Zyczkowski\footnote{Fulbright Fellow.
Permanent address: Instytut Fizyki im. Smoluchowskiego,
 Uniwersytet Jagiello{\'n}ski, ul. Reymonta 4, 30-059 Krak{\'o}w, Poland}
}

\address{Institute for Plasma Research, University of Maryland
         College Park, MD 20742 USA}

\author{Pawe\l{} Horodecki}
\address{Faculty of Applied Physics and Mathematics\\
Technical University of Gda\'nsk, 80--952 Gda\'nsk, Poland}

\author{Anna Sanpera and Maciej Lewenstein}
\address{CEA/DSM/DRECAM/SPAM, Centre d'Etudes de Saclay,\\
\mbox{91191 Gif-Sur-Yvette, France.}}

\date{\today}
\maketitle
\begin{abstract}
The question of how many entangled or, respectively, separable states are there 
in the set of all quantum states is considered. We propose a
natural measure in the space of density matrices $\varrho$ describing 
 $N$--dimensional quantum systems. We prove that
under this measure, the set of separable states possesses
a nonzero volume. Analytical
lower and upper bounds of this volume are also derived  
for $N=2 \times 2$, and $N=2 \times 3 $ cases.
Finally,  numerical Monte Carlo calculations 
allow to estimate the volume of 
separable states providing numerical evidence that it decreases exponentially 
with the dimension of the composite system.
 We have also analyzed a conditional measure of separability
 under the condition of fixed purity.
  Our results display a clear dualism between purity and separability:
entanglement is typical for pure states,
while separability is connected with quantum mixtures. 
In particular, states of sufficiently low purity are necessarily separable.
\end{abstract}

\narrowtext
\newtheorem{definition}{Definition}
\newtheorem{theorem}{Theorem}
\newtheorem{lemma}{Lemma}
\newtheorem{conclusion}{Corollary}

\section{Introduction}

The question of quantum inseparability and entanglement of mixed states
has attracted much attention recently. This problem is, by far, more
complicated than the analogous one for pure states \cite{Werner} an
involves subtle effects like ``hidden nonlocality'' \cite{Popescu}
or ``distillation of entanglement''\cite{Bennet,mydest}. Generally speaking,
one is interested in inseparable states as
the states containing Einstein-Podolsky-Rosen (EPR) correlations.
In fact, all inseparable mixed states have nonzero ``entanglement
of formation'' \cite{Bennett1} which means that to build them
 a nonzero amount
of pure entangled states is needed.
In particular if a source emits pairs of
particles in {\it unknown} pure states, so that they form a
quantum ensemble described by an
inseparable density matrix, then it follows that
the source {\it must} emit  some entangled pairs
with a nonzero probability.
In this  sense the inseparable mixed states can
be vieved as entangled in correspondence to the entangled pure states.

One of the  fundamental questions concerning those subjects is
to estimate how many entangled (disentangled)
states exist among all quantum states.
More precisely,  one can consider  the problem of quantum 
separability--inseparability from the
measure theoretical point of view, and ask
about relative volumes of both sets.
There are three  main reasons of importance of this
problem. The first reason, of some philosophical implications,
may be contained in the questions ``Is the world {\it more classical}
or {\it more quantum}?
Does it contain more quantum correlated (entangled) states than classically correlated ones?''.
The second reason has a more practical origin.
Analyzing some features of entanglement one often has to rely on
numerical simulations.
It is then important to know, to what extend 
entangled quantum states may be considered as typical.
Finally, the third reason has a physical origin. 
Physical meaning of separability has been
recently associated
with the possibility of partial time reversal \cite{tarrach} (see
also \cite{Busch}).
Separable states of
composite systems allow time reversal in one of subsystems, without
loosing their physical relevance. However, for a system of a
dimension $N\ge 8$
the fact that a state admits 
partial time reversal is not sufficient to assure separability,
and  counterexamples have been found\cite{jatran}.
Moreover, it has been recently shown that none of those counterexamples can be distilled to a singlet form \cite{dist}.
Therefore, it seems pertinent to investigate how frequently such peculiar
states appear. At  first glance  it
seems quite likely that such states
form a set of measure zero, and that from a measure theoretical point of
view the set of separable states and the states that admit partial time
reversal have equal volumes.

In this paper we make an attempt to answer at least the two first of
the above formulated questions.
We also give a qualitative argument why the last conjecture
fails.
To this aim we propose a simple and natural measure on the
set $\cal S$ of density matrices acting on a finite dimensional 
Hilbert space $\cal H$. 
Using this measure we estimate the 
relative volume of the set of separable states ${\cal S}_{sep}$.
The upper (lower) bound on this volume is obviously
the lower (upper) bound on the relative volume of the set
of inseparable (entangled) states $S_{insep}=S \setminus S_{sep}$.

The paper is organized as follows. 
Section II contains our definition of the natural measure in $\cal S$. In Section III  we recall basic definitions of separable states, and 
prove that for any compound system ${\cal S}$, the volume of ${\cal S}_{sep}$
is nonzero regardless the number of subsystems  it contains and its (finite) dimension.
 This is achieved by proving the existence of a  topological lower bound of this volume.
Better lower bounds are also calculated analytically by
analyzing the relation between the purity of the state and its separability.
 In Section IV analytic upper bounds on the volume of ${\cal S}_{sep}$ are found. 
The study of inseparable states with positive partial
 transposition is presented in Section V. In Section VI 
we present the  estimates on the volume of separable states obtained
 by the Monte Carlo numerical simulations. This section is self--contained,
 and includes also a simplified corollary of the results of Sections 
II--V, and a  discussion of the dualism between purity and separability.
 We conjecture that the volume of separable states decreases
exponentially with the dimension of the Hilbert space. Finally,
 Section VII contains our conclusions and open questions.

The reader should note the that the Sections II--V have a rather formal
 mathematical character. The results of these Sections provide a rigorous 
base for the numerical calculations of Section VI, but the detailed knowledge
of the proofs is by no means necessary to understand the main message of the paper.
 The reader who is not interested in such rigorous  
proofs of the presented results, may well skip Sections  III--V,
and go straight to the Sections VI and VII.

\section{Natural measure of the set of quantum states}

Let us consider a set of states in a $N$--dimensional Hilbert space
${\cal H}$.  In particular, $\cal H$ may describe 
 a composite system with
$m$ component subsystems: 
${\cal H}=\mathop{\otimes}\limits_{i=1}^{m} {\cal H}_i$, where
$ \prod_{i=1}^{m} N_i=N$.

An operator $\varrho$ acting on $\cal H$ describes a {\it state} if $\text{{\rm Tr}}\varrho=1$
and if $\varrho$ is a positive operator, i.e.
\begin{equation}
\text{{\rm Tr}}( \varrho P)\geq0
\end{equation}
for any projector $P$.
Any state  represented by a
density matrix $\varrho$ can in turn  be represented by its spectral decomposition:
\begin{equation}
\varrho=\sum_{n=1}^{N} \Lambda_n P_n, \ \ \sum_{n=1}^{N} \Lambda_n=1,\ \  \Lambda_n \geq 0,
\end{equation}
where $P_n$ form a complete set of orthogonal projectors.
Thus the set of states can be viewed as a Cartesian product of sets:
\begin{equation}
{\cal S}={\cal P} \times \triangle.
\label{tri}
\end{equation}
The set ${\cal P}$ denotes the family of complete sets of
orthonormal  projectors  $\{P_i\}{}_{i=1}^{N}$,
$\sum_{n=1}^{N} P_n =I$, where $I$ is the identity matrix.
There exist the unique,  uniform measure $\nu$ on $\cal P$ induced by
the Haar measure on the group of unitary matrices $U(N)$.
Integration over the set $\cal P$ amounts thus to the integration of
the corresponding angles and phases in $N$-dimensional complex space
that determine the families of orthonormal projectors (or, alternatively 
speaking the unitary matrix that diagonalizes $\varrho$).

The symbol 
$\triangle $ in Eq. (\ref{tri}) represents there the 
set of all $\Lambda_n$'s, which is a subset of the
$N-1$ dimensional linear
submanifold of real space ${\cal R}^{N}$, defined by the trace condition, 
$ \sum_{n=1}^{N} \Lambda_n=1$. Geometrically, $\triangle $ 
is defined as a convex hull (i.e. a set of all convex combinations
of the edge points) $\triangle = {\rm conv}\{ {\bf x_i} \in {\cal R^N}:
{\bf x_i}=(0, ... , \mathop{1}_{i} , ... , 0), i=1, ... , N  \}$.
Since the simplex $\triangle $ 
is a subset of the $N-1$ dimensional hyperplane, there exist
a natural  measure on $\triangle $ which is defined as a usual normalized
Lebesgue measure ${\cal L}_{N-1}$ on ${\cal R}^{N-1}$.
More specifically, any measurable function $f(.)$ of
 $\Lambda_1,\ldots,\Lambda_N$ can be integrated with this measure:
\begin{eqnarray}
& &\frac{1}{\rm{V_s}}\int_0^1 \! d\Lambda_1\ldots \! 
 \int_0^1 d\Lambda_Nf(\Lambda_1,\ldots,\Lambda_N)\delta(\sum_1^N
\Lambda_n-1)=\nonumber \\
& &\frac{1}{\rm{V_s}}\int_0^1 \! d\Lambda_1\ldots \!
  \int_0^1 d\Lambda_{N-1}f(\Lambda_1,\ldots,\Lambda_{N-1},1
-\sum_0^{N-1}\Lambda_n),
\end{eqnarray}
where the normalization constant $\rm{V_s}$ equals
 to the volume of the set  $\triangle$ in ${\cal R}^{N-1}$,
 whereas $\delta(.)$ denotes the Dirac's delta distribution.
The two above discussed measures induce a natural measure on
${\cal S}$:
\begin{equation}
\mu = \nu \times {\cal L}_{N-1}.
\end{equation}

\section{Volume of the set of separable states}

\subsection{Preliminaries - separable states}

Throughout this paper we shall assume that the Hilbert space of
the considered quantum system has an arbitrary but finite dimension.
To make  further considerations more clear, we start from the following
notation and definitions.
Recall first that the space ${\cal A}$ of operators acting on ${\cal H}$
constitute a new Hilbert space (a so called Hilbert-Schmidt space)
with the scalar product:
\begin{equation}
\langle A,B\rangle=\text{{\rm Tr}}(B^\dagger A).
\label{il}
\end{equation}
It induces a natural norm (a trace norm):
\begin{equation}
||A||=\sqrt{\text{{\rm Tr}}(A^\dagger A)}.
\end{equation}
which, according to the condition ${\rm dim} {\cal H} < \infty$
is topologically equivalent to all other norms on ${\cal A}$, in
particular to the norm
$||A||'=\text{{\rm Tr}}|A|$.
Furthermore, let us  recall that:
\begin{definition}
The state $\varrho$
acting on the Hilbert space ${\cal H}={\cal H}_1 \otimes {\cal H}_2$
is called separable
\footnote{The presented definition of separable states is due to Werner
\cite{Werner}, who called them classically correlated states.}
if it can be approximated in the trace norm by the states of the form
\begin{equation}
\varrho=\sum_{i=1}^kp_i\varrho_i\otimes\tilde \varrho_i
\label{sep}
\end{equation}
where $\varrho_i$ and $\tilde\varrho_i$ are states on ${\cal H}_1$ and
${\cal H}_2$ respectively.
\end{definition}
Usually one deals with a finite dimensional Hilbert space dim${\cal H}=N$.
For this case it has been shown \cite{jatran} 
that any separable state can be written as a convex combination of {\it finite}
 product pure states, i.e. in those cases the ``approximation'' part
 of the definition is  redundant.

It has also been shown \cite{Peres} that the necessary condition for separability
of the state $\varrho$ is positivity of its
partial transposition $\varrho^{T_2}$.
The latter is
defined in an arbitrary orthonormal product basis $|f_i\rangle
\otimes |f_j\rangle$ as a  matrix with elements:
\begin{equation}
\varrho^{T_2}_{m m',n n'}\equiv
\langle f_m|\otimes \langle f_{m'} | \varrho^{T_2}| f_n \rangle\otimes |f_{n'} \rangle=
\varrho_{m n',n m'}.
\label{tr}
\end{equation}
Although the matrix $\varrho^{T_2}$ depends on
the used basis, its eigenvalues do not.
Consequently, for any state the above condition can be checked
using {\it an arbitrary} product orthonormal 
basis\footnote{As the full transposition of a positive operator is also
positive, positivity of the partial transposition  $\varrho^{T_2}$
is equivalent to positivity of 
the partial transposition  $\varrho^{T_1}$ (defined in analogous way).}.
For systems of dimensions $2 \times 2$
and $2 \times 3$
the partial transposition condition is also a sufficient one
\cite{my1} and thus the set of separable states
is  completely characterized by this condition.

The definition of separable states can be easily generalized
to systems composed of more than two subsystems:
\begin{definition}
The state $\varrho$
acting on the Hilbert space ${\cal H}=\mathop{\otimes}
\limits_{l=1}^{m} {\cal H}_l$
is called separable
if it can be approximated in the trace norm by the states of the form
\begin{equation}
\varrho=\sum_{i=1}^k p_i \mathop{\otimes}\limits_{l=1}^{m}
\varrho^{l}_i
\label{sep2}
\end{equation}
where $\varrho^{l}_i$ are states on ${\cal H}_l$ .
\end{definition}
Straightforward generalization of the proof about decomposition
from Ref. \cite{jatran} gives us the possibility of omitting the
approximation part in the definition:
\begin{lemma}
Any separable state $\varrho$ of a system composed by $m$
subsystems can be written as:
\begin{equation}
\varrho=\sum_{i=1}^k p_i P^{i}_{prod}, \  k \leq N^2
\end{equation}
where $P^{i}_{prod}$ are  pure product states
having  the m-decomposable form
$\mathop{\otimes}\limits_{l=1}^{m} P_l$, where $P_l$ are projectors acting on ${\cal H}_l$.
\end{lemma}
It is worth mentioning that minimal decompositions
 with $k=N$ can be always found for $N=4$ \cite{tarrach,wooters}.

\subsection{Existence of nonzero lower bound for the volume 
of separable states}

We shall prove now that the volume of the set of
separable states is nonzero independently of the dimension
of the Hilbert space and the number of subsystems $m$ composing it.

\noindent For our purposes we first prove the simple lemma:
\begin{lemma}
If the hermitian operator $A \in {\cal A}$ satisfies
$\langle A, \mathop{\otimes}\limits_{i=1}^{m} P_i\rangle=0 $
for any product projectors $\mathop{\otimes}\limits_{i=1}^{m} P_i$, then
it is a trivial zero one.
\end{lemma}

{\it Proof.-}
Let us consider an arbitrary orthogonal (in the sense
of scalar product (\ref{il})) product hermitian basis
in the space of operators ${\cal A}$, i.e. a basis such that 
any of its elements  is a product of hermitian matrices
(for instance, in the $2 \times 2$ case the basis could consist of
products of Pauli matrices
$\sigma_{n}\otimes\sigma_{m}$, with $n, m = 0, 1, 2, 3$;
$\sigma_{0}=I$ ).
Hermicity assures that any element of the basis
can be written as a real combination of product projectors.
Any coefficient of the expansion of $A$ in this basis
is given by the scalar product (\ref{il}) of $A$ and the corresponding 
basis element. From
the general assumption of vanishing of formulas of type
$\langle A, \mathop{\otimes}\limits_{i=1}^{m} P_i\rangle=0 $,
we obtain immediately that all expansion coefficients must vanish. Hence $A$
must be equal to the zero operator.

We can now  propose the following theorem :
\begin{theorem}
Let $\triangle_{\epsilon}$ be a simplex defined as
$\triangle_{\epsilon} = {\rm conv}\{{\bf y_i} \in {\cal R}^N: {\bf y_i}=\epsilon{\bf x_i} +
(1-\epsilon){\bf z_{I}}; i=1, ... , N; \
{\bf z_{I}}= ({1 \over N}, ... , {1 \over N})\}$.
Let define a set
${\cal Q}_{\epsilon} = {\cal P} \times \triangle_{\epsilon}$.
Then there exists some positive $\epsilon$ such that
 ${\cal Q}_{\epsilon} \subset {\cal S}_{sep}$, where ${\cal S}_{sep}$
 represents the set of separable states.
\end{theorem}

The meaning of the above theorem is straightforward. It says that all states 
in the sufficiently small neighborhood of the maximally mixed  state
$\varrho_{I}={1 \over N} I$ (which
is represented in $\triangle $ as a
point ${\bf z_I}=({1 \over N}, ... , {1 \over N})$ for any
 chosen spectral decomposition of unity) are necessarily separable.
 Note that by definition, the simplex $\triangle_{\epsilon}$ has edges  $\epsilon$ times
smaller than $\triangle$, so that its volume $\mu(\triangle_{\epsilon})=
\epsilon^{N-1}\mu(\triangle)=\epsilon^{N-1}$,
 since according to our normalization $\mu(\triangle)=1$.

{\it Proof .-}
Suppose, conversely that for any positive $\epsilon$
the set ${\cal Q}_{\epsilon}$ contains some inseparable state
$\varrho_{insep}$. It is easy to see that then there must exist
a sequence of inseparable states $\varrho^{n}_{insep}$ convergent to the
maximally mixed state $\varrho_{I}$.
According to Lemma 1  and Theorem 1 from
Ref. \cite{my1} there exist a
sequence of operators $A_n$ separating the states $\varrho^{n}_{insep}$
from the state $\varrho_{I}$ in the sense that for any $n$ it holds:
$\langle A_n, \varrho_{insep}^n \rangle<0 $ and
$\langle A_n, \varrho_{I} \rangle\geq 0 $.
Moreover, from the quoted results it follows that
$\langle A_n ,\sigma \rangle\geq 0 $ for any $\sigma \in {\cal S}_{sep}$.
Let us normalize the operators $A_n$ by 
introducing $\tilde{A_n}=A_n/||A_n||$
( $||A_n||=\sqrt{\langle A^{\dagger}A \rangle}$).
These operators satisfy:
\begin{equation}
\langle \tilde{A}_n, \varrho_{insep}^n \rangle<0,
\langle \tilde{A}_n, \sigma \rangle\geq 0 \ \ \mbox{ for any} \ \
\sigma \in {\cal S}_{sep}.
\label{odd}
\end{equation}
In particular it holds $\langle \tilde{A}_n, \varrho_{I} \rangle\geq 0$.
>From construction the sequence $\tilde{A}_n$
belongs to the sphere in the finite dimensional space
${\cal A}$. As the latter is a compact set, the sequence includes
some subsequence  $\tilde{A}_{n(k)}$ which is convergent to
some nonzero operator $\tilde{A}_*$ ($||\tilde{A}_*||=1$).
>From (\ref{odd}) and continuity
of scalar product it follows that the limit operator also satisfies:
\begin{eqnarray}
\langle \tilde{A}_*, \sigma \rangle\geq 0 \ \ \mbox{ for any} \ \
\sigma \in {\cal S}_{sep}\,\,.
\label{dod}
\end{eqnarray}
Now using (\ref{odd}) and the Schwarz inequality we obtain:
\begin{eqnarray}
0\leq& & \langle \tilde{A}_{n(k)}, \varrho_{I} \rangle
 = \langle \tilde{A}_{n(k)}, \varrho_{I} - \varrho^{n(k)}_{insep}
\rangle
+ \langle \tilde{A}_{n(k)}, \varrho^{n(k)}_{insep} \rangle \nonumber \\
& &\leq \langle \tilde{A}_{n(k)},\varrho_{I} - \varrho^{n(k)}_{insep}  \rangle
 \leq ||\tilde{A}_{n(k)}|||| \varrho_{I} - \varrho^{n(k)}_{insep} ||\nonumber\\
& &=|| \varrho_{I} - \varrho^{n(k)}_{insep}||.
\end{eqnarray}
Taking the limit with respect to  $k$ we obtain:
\begin{equation}
\text{{\rm Tr}}\tilde{A}_*=\langle \tilde{A}_*, \varrho_{I} \rangle=
\lim_{k \rightarrow \infty} 
\langle \tilde{A}_{n(k)}, \varrho_{I} \rangle=0\, .
\label{zero1}
\end{equation}
Hence $\tilde A_*$ is traceless, which is in contradiction
with (\ref{dod}).
Indeed, if the operator $\tilde{A}_*$ is to be nontrivial
(the construction implies its unit norm) then
there must exist some product state
$P_{prod}\equiv\mathop{\otimes}\limits_{i=1}^{m} P^{i}$
such that $\langle \tilde A _*, P_{prod} \rangle \neq 0 $ (see lemma 2).
Since, on the other hand,  one requires
the trace of $\tilde A_*$ to vanish, one obtains that
$\langle \tilde A_*, P_{prod} \rangle =
- \langle \tilde A_*, I - P_{prod}\rangle $.
Hence, one of the separable states $\sigma'=P_{prod} $, $\sigma''=
{1 \over N - 1}(I - P_{prod})$ violates the condition (\ref{dod}),
which gives the  expected contradiction.
The above theorem leads immediately to the following one:

\begin{theorem}
The measure $\mu({\cal S}_{sep})$ of separable states is a nonzero
one. In particular there exists always some $\epsilon > 0$ such that
the following inequality holds
\begin{equation}
\mu ({\cal S}_{sep}) \geq \mu (\triangle_{\epsilon}) = {\epsilon}^{N - 1} > 0.
\end{equation}
\end{theorem}

\noindent As an illustration of the above theorem, 
 let us consider the $2 \times 2$ or
$2 \times 3$ cases ($N=4, 6$) for which  separability 
is equivalent to the positivity of the partial transposition.
It is easy to see that the spectrum of the partially transposed
density matrix must belong to the interval $[-{1 \over 2}, 1]$.
Hence any state of the form
$\varrho=(1-p){I \over N} + p\tilde{\varrho}$, for an arbitrary
$\tilde\varrho$ and $p\leq {2/(2+N)}$, has a positively defined partial transposition, and thus  is separable for the considered
cases. As the maximal value of $p$ is 1/3, or 1/4,  it means that
the value of $\epsilon$ in the above theorem can be
estimated just by ${1 \over 3}$, or ${1 \over 4}$  for $N=4, 6$, respectively.
In the next section we shall show that those
bounds can be significantly improved.

\subsection{Purity and separability}

As we have shown, all states in the small enough
neighborhood of the totally mixed state $\varrho_{I}=I/N$
are separable. On the other hand, we know that in the
subspace of all pure states, the measure
of separable states is equal to zero \cite{Popescu}.
It is, therefore, interesting to investigate the relationship 
between entanglement and  mixture of quantum states.
 A qualitative characterization of the degree of  mixture is provided by
the von Neumann entropy
$H_1(\varrho)=-{\rm {\rm Tr}}(\varrho \ln \varrho)$.
 Another quantity, called participation ratio: 
\begin{equation}
R(\varrho)={1\over \rm{{\rm Tr}}( \varrho^2)}
\end{equation}
is often more convenient for calculations. It varies from the
unity
(for pure states) to $N$ (the totally mixed state $\varrho_{I}$) and may be
interpreted
as an effective number of states in the mixture. This quantity,
applied in solid state physics long time ago \cite{Weaire},
is related to the von Neumann-Renyi entropy of order two,
$H_2(\varrho)=\ln R(\varrho)$.
The latter, called also purity of the state,
together with  other quantum Renyi entropies
$H_q(\varrho)=(\ln [$Tr$ \varrho^q])/(1-q)$  
 is used,  for $q\ne 1$, as a measure of how much
a given state is mixed. It  has also been applied for the derivation
of some necessary conditions of separability in Ref. \cite{alfa}.
In Section VI we shall demonstrate, using numerical simulations,  that the
participation ratio (and other von Neumann--Renyi entropies) 
allows  to establish a dualism between purity and separability of the states
of composite systems.
In this subsection we use it to calculate a
natural lower bound on the volume of separable states
for dimensions $N=4, 6$.

\noindent For this purpose consider the following lemma:
\begin{lemma}
If the state $\varrho$ satisfies
\begin{equation}
R(\varrho)\geq N - 1
\label{cond}
\end{equation}
where $N$ is the dimension of  ${\cal H}$, then $\varrho^{T_{2}}$ is positive
 defined, i.e. its spectrum  $s( \varrho^{T_{2}} )$ belongs to the simplex $\triangle$.
\end{lemma}

{\it Proof.-} Let us denote by $B_{N}(r,P)$ 
the ball in the space $R^{N}$ with radius $r$ and center $P$ and by $S_{N}(r,P)$ its
surface. The
condition (\ref{cond}) is invariant with respect to the partial
transposition, because ${\rm {\rm Tr}}(\varrho^2)= {\rm {\rm Tr}}(( \varrho^{T_2})^2)$.
That implies that
$s( \varrho^{T_2})\in B_{N}(r,{\bf z_{I}})$ with
$r={1 \over \sqrt{N-1}},{\bf z_{I}}=({1 \over N}, ... ,{1 \over N})$.
Let define the $(N-1)$-dimensional linear 
 manifold ${\cal M}_{N-1}=\{ {\bf x}=( x_1, ..., x_N ),
\sum^{N}_{i=1} x_i=1\}$.
We only need to show that its intersection with the ball is included
in the simplex $\triangle $,  i.e. that  the new $(N-1)$-dimensional ball
$B'_{N-1} \equiv B_{N}(r,{\bf z_{I}})
\cap {\cal M}_{N-1} \subset \triangle $. This can be seen 
in the following way.  It follows from
the high symmetry of the sphere and the invariance of the simplex
under cyclic permutations of coordinates, that the center of this intersection
is again ${\bf z_{I}}$. Hence the radius $r'$ of
$B_{N-1}$
can be calculated immediately by taking the distance from
an arbitrary point from the surface
$S_{N}(r,{\bf z_{I}})
\cap {\cal M}_{N-1}$ (say, for example from the point 
$(0, {1 \over N-1}, ...,{1 \over N-1}$)
to the point ${\bf z_{I}}$). It is elementary to show that $r'={1 /\sqrt{N(N-1)}} $.
On the  other hand, one can calculate the
maximal radius $r''$ of the ball of the type
$B''_{N-1}(r'',{\bf z_{I}})$ included in $\triangle$ by calculating the
minimal distance of ${\bf z_{I}}$ to the boundary of $\triangle$. 
To this aim we have to minimize $(r'')^2=\sum_{i=1}^N (x_i-1/N)^2$ with
 the constraints $\sum_{i=1}^N x_i=0$, and $x_0=0$. 
Using Lagrange
multipliers we obtain immediately $r''=r'$, and hence $B_{N-1}$ belongs to $\triangle$,
which ends the proof.

Now using the explicit expressions for the volume of a $(N-1)$-dimensional ball
($V_{N}(r)={ \pi^{N - 1 \over 2} r^{N-1} / \Gamma( {N - 1 \over 2})}$),
and for the volume of the simplex $\triangle$ belonging to the manifold
${\cal M}_{N-1}$ ($V_{\triangle}={\sqrt{N} / (N-1)!}$),
one  can obtain the lower bound of the volume of states with positive
partial transposition,
\begin{equation}
\tau_{N}={(N-1)! \pi^{N-1 \over 2} \over N^{N \over 2} (N-1)^{N-1 \over 2}
\Gamma({N + 1 \over 2})}.
\label{eq18}
\end{equation}
Recalling that for  Hilbert spaces of dimensions $N=2 \times 2$ and $N=2 \times 3$ the states with positive
partial transposition are the separable states,  Eq. (\ref{eq18}) leads directly to
the following theorem:
\begin{theorem}
If the participation ratio satisfies
$R(\varrho)\geq 3$ $(R(\varrho)\geq 5)$ for $N=4$ $(N=6)$ then the
state $\varrho $ is separable.
\end{theorem}

\noindent Therefore, the measure $\mu({\cal S}_{sep})$ of separable states 
is restricted from below by
the inequality $\mu ({\cal S}_{sep}) \geq {\pi \over 6 \sqrt{3}}\simeq 0.302$ for
$N=4$ and  ${8 \pi^2 \over 625 \sqrt{5}}\simeq 0.056$ for $N=6$.

\section{Upper bounds on the volume of separable states}

In this section we seek for upper bounds on the volume of $V_{sep}$, 
or, equivalently,  lower bounds on the set of inseparable states $V_{insep}$. 
Several necessary conditions for separability have been recently established with the
aid of positive maps. We should use them to determine an upper bound on $V_{sep}$.
As we shall see, these conditions are in some way complementary and can be combined to
obtain a better estimate of the upper bound of $V_{sep}$.  

\noindent Our first estimate relies on the positivity of the partial transposition. 
It is valid for any dimension, but we shall apply it 
to composite systems of dimension 2$\times $2. 
Note that if a state has a partial transposition which 
is not positively defined, then the state is necessarily inseparable.
 Before proceeding further we should first recall the Schmidt decomposition of a pure
 state $|\Psi\rangle\in {\cal H}={\cal H}_1 \otimes {\cal H}_2$,
 ${\rm dim}{\cal H}_1=N_1$, ${\rm dim}{\cal H}_2=N_2$, $N_1\times N_2=N$,
\begin{equation}
|\Psi\rangle=\sum_{i=1}^{{\rm min}(N_1,N_2)}a_i |e_i\rangle\otimes|f_i\rangle,
\label{state}
\end{equation}
where  $|e_i\rangle\otimes|f_i\rangle$ form a bi-orthogonal basis 
$\langle e_i | e_j\rangle =\langle f_i | f_j\rangle=\delta_{ij}$, and  
$0\le a_i\le 1$ denote the coefficients of the Schmidt decomposition 
with the condition $\sum_i a_i^2=1$.
It is straightforward to see 
that $P_{\Psi}^{T_2}=(|\Psi\rangle\langle \Psi|)^{T_2}$ 
has eigenvalues $a_i^2$ for $i=1,\ldots,{\rm {min} (N_1,N_2)}$ and
$\pm a_ia_j$ for $i\ne  j$.

\noindent We can now state our first lemma:
\begin{lemma}
If in the range of a state $\varrho$ there exists $|\Psi\rangle$ such  that
\begin{equation}
\Lambda= \langle \Psi|\varrho^{-1}|\Psi\rangle^{-1}  > {1 \over 1 +
 \max_{ i\neq j}(a_ia_j)},
\label{lem4}
\end{equation}
then $\varrho$ is inseparable.
\label{lema4}
\end{lemma}

{\it Proof.-}
According to Ref. \cite{MS}, any state $\varrho$ can be expressed as:
\begin{equation}
\varrho=\Lambda P_{\Psi}+ (1-\Lambda)\tilde \varrho,
\end{equation}
where $P_{\Psi}$ is a projector onto $|\Psi\rangle$ and 
$\tilde\varrho$ is a (positively defined) state. Thus:
\begin{equation}
\varrho^{T_2}=\Lambda P_{\Psi}^{T_2}+ (1-\Lambda)\tilde \varrho^{T_2}.
\end{equation}
Recall that for any $N_1$, $N_2$, the eigenvalues of $\tilde\varrho^{T_2}$ belong to
 the interval $[-1/2,1] $. Let $|\Psi_{neg}\rangle$ denote the eigenvector
 corresponding to the minimal eigenvalue of $P_{\Psi}^{T_2}$:
 $-{\rm max}_{ i\neq j}(a_ia_j)$. We have thus
\begin{equation}
\langle\Psi_{neg}|\varrho^{T_2}|\Psi_{neg}\rangle 
\le -\Lambda ({\max}_{ i\neq j}(a_ia_j)) + 1-\Lambda<0,
\end{equation}
because $\langle\Psi_{neg}|\tilde\varrho^{T_2}|\Psi_{neg}\rangle \le 1$. 
The above inequality implies that $\varrho^{T_2}$ is not positively defined when
 condition (\ref{lem4}) holds,  
and therefore $\varrho$ is not separable. Note, that the lemma 
(\ref{lema4}) can be in particular applied to the eigenvectors of $\varrho$:

\begin{lemma}
If $\varrho$ has an eigenvector $|\Psi\rangle$ corresponding  to the eigenvalue
$\Lambda$ such that the condition (\ref{lem4}) holds, 
then $\varrho$ is inseparable.
\label{lema5}
\end{lemma}
The eigenvalue $\Lambda$
can fulfill
the above condition if and only if it is the largest eigenvalue, because
it must be larger than $2/3$. The corresponding normalized
eigenvector, however, is absolutely arbitrary, and  according to
invariant measure on the group it can be
generated simply by a
uniform probability on the $N$ dimensional unit sphere.
 This implies, as we shall see below,
 that the Schmidt coefficients $a_i$ are also absolutely arbitrary
and distributed uniformly on the octant of the $\sqrt N$-dimensional sphere.

Consider the  $N_1=N_2=K$ case.
Any vector in $N=K^{2}$--dimensional space from the unit sphere can be represented
by a row of complex numbers $x_{i}$
with the condition $\sum^{N}_{i}|x_{i}^{2}| = 1$.
In any product basis, we can view it as a $K\times K$ matrix $C_{ij}$ with
$i,j=1,\ldots,K$, and  with the condition ${\rm Tr}( C^{\dagger}C) ) = 1$.
We seek the uniform distribution on the set of such matrices.
But from the polar decomposition theorem any
matrix of such a type can be represented in the form:
\begin{equation}
C=U'DU
\end{equation}
where $U'$ and $U$  are some unitary matrices, while $D$ is
diagonal matrix with nonnegative elements (eigenvalues). These 
eigenvalues are nothing else   but $a_i$.
The reason is that the above form, which
 is the analog of the spectral decomposition
of the hermitian matrix, is at the same time the Schmidt 
decomposition written in the matrix notation. In our case (taking into
 account the above mentioned trace condition), the spectrum of $D$
is represented by the point belonging to the octant area
of the sphere.  These  leads  to the measure
\begin{equation}
\mu'(\Psi)=\nu( U'(K) ) \times \nu( U(K) ) \times \mu (D),
\label{mu'}
\end{equation}
where the first two measures are Haar measures on the unitary
 group $U(K)$, and the last one is the uniform (Lebesgue) measure on the 
octant of the ball in $K$ dimensional space. Similar results can be 
straightforwardly generalized for the cases $N_1\neq N_2$.

If one calculates now  the measure (\ref{mu'}) for $|\Psi\rangle$,
 and combines it with the uniform measure on the simplex
$\mu_{\triangle}$,
one could estimate an upper bound of separable states:
\begin{eqnarray}
& &\mu(S_{sep})\le 1 -\nonumber\\
 & &\int  \Theta(\max_i\Lambda_i- ( 1 +
{\max}_{ i\neq j}(a_ia_j))^{-1})\,d\mu '(\Psi) d\mu_{\triangle},
\end{eqnarray}
where $\Theta$ denotes the Heavyside function. 
The double integration over the unitary groups that
contains $\mu'(\Psi)$ can be easily performed, since neither 
$\Lambda_i$ nor $a_i$ depend on the direction of $|\Psi\rangle$.
This is the first qualitative argument that the measure
of inseparable states does not vanish.

Moreover, recently \cite{maps} a new separability condition
have been introduced with the aid of positive maps condition
\cite{my1}:
if the state $\varrho$ is separable then
$I \otimes \varrho_1 - \varrho$ must be positive, where 
$\varrho_1$ is the reduced density matrix.
It implies for any $|\Psi\rangle$ that
${\rm {\rm Tr}}((I\otimes\varrho_1)P_{\Psi}) \geq {\rm {\rm Tr}}(\varrho P_{\Psi})$.
Straightforward estimation tells us that for any separable
state it must hold that $\langle \Psi|\varrho|\Psi \rangle
\leq\mathop{\rm{max}}\limits_{i } a_i^2$, where   $a_i$
are again the Schmidt decomposition coefficients of $\Psi$. That implies 
a lemma analogous to lemma (\ref{lema4}).

\begin{lemma}
If in the range of a state $\varrho$ there exists $|\Psi\rangle$ such  that
\begin{equation}
\Lambda= \langle \Psi|\varrho|\Psi\rangle > \max\limits_{i } ~ a_i^2,
\label{lem6}
\end{equation}
then $\varrho$ is inseparable.
\label{lema6}
\end{lemma}

This lemma is neither stronger nor weaker than the lemma  (\ref{lema5}).
 If we apply it  to eigenvectors of $\varrho$,
however, the relevant 
eigenvalue need not be the maximal. In the case $2\times 2$ we can combine both
conditions (Lemma \ref{lema5}--\ref{lema6}) 
to obtain a better estimate on the upper bound of $\mu(\rm{Sep})$
\begin{eqnarray}
1-\mu(S_{sep})\ge&& \frac{4}{V_{\Delta}V_{oct}}\int_0^1 d\Lambda_1 \int_0^{1-\Lambda_1}
 d\Lambda_2 \int_0^{1-\Lambda_1-\Lambda_2} d\Lambda_3\nonumber \\
& &\int_{a_1\ge 0}da_1\int_{a_2\ge 0}da_2\Theta(\Lambda_1-(1+a_1a_2)^{-1})\nonumber\\
& &\Theta(\Lambda_1-
{\rm max}(a_1^2, a_2^2)) \delta(a_1^2+a_2^2-1),
\end{eqnarray}
Notice that in the above expression the first three integrals
 are over the eigenvalues of $\varrho$ that are located in the simplex $\Delta$,
 whereas the remaining two integrals are on the eigenvalues of $D$ from 
the octant area of the sphere. The integrals
 can be calculated analytically but the resulting expressions are very complex. 
After a tedious, but straightforward calculation we obtain: 
\begin{equation}
\mu(S_{sep})\leq 0.863
\end{equation}

In general for an arbitrary dimension $N=N_1\times N_2$ and $K={\rm min}(N_1,N_2)$; $1-\mu(S_{sep})$ can be 
estimated from below by a bound $b(N_1,N_2)$ using the above method.
This bound, on the other hand, can be estimated from above by the volume of
 the ``corners'' of the simplex $\Delta$ of the 
sides $1-{1\over K}$ 
regardless the uniform measure on pure states by putting it equal to  unity there.
This follows from
the fact that the condition (\ref{lem6}) can only then be fulfilled, when
an eigenvalue of $\varrho$ is 
larger than $ \max\limits_{i }  a_i^2\ge 1/K$.
The relative volume of such corners
equals $N(1 - {1\over K})^{N-1}$.
Keeping in mind the formula (\ref{state})
the above simple estimation leads to the corollary:
\begin{conclusion}
Consider a quantum system $\in {\cal H}_1 \otimes {\cal H}_2$, where
$dim{\cal H}_1=N_1$, dim${\cal H}_2=N_2$, $N=N_1\times N_2$ and
$K = {\min}( N_1 , N_2 )$.
Then using Lemae 4-6 the volume of separable states is restricted from above by:
\begin{equation}
\mu (S_{sep})\leq 1 - b(N_1,N_2) \nonumber
\end{equation}
where:
\begin{equation}
b(N_1,N_2)\leq (1-{1 \over K})^{N-1}
\label{volN}
\end{equation}
\end{conclusion}

The volume (\ref{volN}),  however, converges asymptotically to the
value 1 as $N_1,N_2$ grow, so that in the limit of large $N$ we get a trivial result
 $\mu(S_{sep})< 1$. At the same time, the numerical results which we
shall present subsequently strongly suggest that there
should exist an upper bound for $\mu(S_{sep})$ converging to zero.
So far, the rigorous proof that in the limit
of higher dimensions  $ \mu(S_{sep}) \rightarrow 0$ remains an open problem.

\section{Inseparable states with positive partial transposition}

As it was mentioned in the introduction for $N \ge 8$ there are states which
are inseparable but have positive partial transposition \cite{jatran,my1}.
Moreover, it has been recently shown that all states of such type
represent ``bound" entanglement in the sense that they cannot
be distilled to the singlet form \cite{dist}.
The immediate question that arises is how frequently such peculiar
states appear in the set of all the states of a given composite
system. This question is related to the role of time reversal in 
the context of entanglement of mixed states \cite{tarrach,Busch}
 Below we provide a qualitative argument that the volume of the set of those states
 is also nonzero:
\begin{lemma}
For $N \ge 8$ the set of inseparable states with positive
partial transposition includes a nonempty ball.
\end{lemma}

{\it Proof.-} Consider the
two sets of quantum states for some composite system:
the set of separable states
$S_{sep}$ and the set of states with positive partial transposition
$T$. The first of them is convex and compact.
The second one is convex set \footnote{In fact it is also
compact but we shall not need this property here.}.
Since positivity of partial transposition is
necessary for separability,  we have obviously $S_{sep} \in T$.
Consider any state $\sigma$ belonging to
$T$ but not to $S_{sep}$ (we know that for $N\ge 8$ such states exist).
Let us take the ball $B(r,\varrho_{I})$
around the maximally chaotic state $\varrho_{I}$ such that
the whole $B(r,\varrho_{I})$ belongs to $S_{sep}$
(the ball can be in principle defined in an arbitrary norm as all
norms are topologically equivalent, c.f. section III).
Obviously such a ball exists; otherwise $\varrho_{I}$
would belong to the boundary of $S_{sep}$ .
It can be shown that the latter would contradict
the fact of a nonzero volume (see section III).
Consider the sequence of balls obtained from $B(r,\varrho_{I})$
by a translation of the center $r$, and a rescaling:
 $B_n= B({r\over n}, (1 -{1 \over n})\sigma +{1 \over n}\varrho_{I})$. 
Some of $B_n$'s have to include no separable states. Otherwise, if any $B_n$
 included some separable states e.g. $\varrho^{sep}_{n}$,
by virtue of compactivity of $S_{sep}$,
the state $\sigma$ would be separable as the limit of
a sequence of separable states $\varrho^{sep}_{n}$.
Hence, some $B_{n_{0}}$ do not belong to set of separable states.
But on the other hand, 
any state from $B_{n_0}$ is a convex combination
of elements of $T$. Thus the whole ball $B_{n_0}$ belongs to $T$,
as the latter is a convex set.
In this way  we have shown that there is some ball $B_{n_0}\subset T$ and
 does not intersect with $S_{sep}$.
But the inseparable states with
positive partial transposition are just the ones belonging to
$T$ and not to $S_{sep}$. This ends the proof of the lemma.

\section{Numerical results}

In this section we provide rather precise estimates of the measure
$ \mu(S_{sep})$ for $N=4, 6$, as well as upper bounds of such measure for
$N>6$. Our results are obtained numerically. This section is self--contained,
 in the sense that can be read directly without passing though the more technical
 sections III-V. It also includes the results obtained in these previous sections.

\subsection{Estimation of volume of separable states}

Our goal is to estimate the volume of the set of separable states
$\mu (S_{sep})$.
For simplicity we discuss states consisting of $2$
subsystems.
Any state describing a mixture of $N_1$ and $N_2$--dimensional
subspaces may be represented by a positive defined  $(N_1N_2) \times (N_1 N_2)$
 hermitian matrix $\varrho$ with trace equal to unity.
 $\varrho^{T_2}$ denotes, as before,  the matrix  partially
 transposed with respect to the second subsystem. As mentioned previously, 
if  $\varrho$ is separable,  then necessarily  $\varrho^{T_2}$ is positive \cite{Peres}.
Moreover, for the simplest $2 \times 2$ and $2 \times 3$
problems this condition is also a sufficient one \cite{my1},
which is not true in the general case $N\ge 8$ \cite{jatran,my1}.
Therefore, the set of separable states
is {\it a subset} of states with positive partial
transpositions. Thus, in order to estimate the volume
of separable states from above, it is sufficient
to find the volume of  the set of states with positive partial transpositions.

Let us remind that according to the section II any state (any density matrix)
can be represented 
in a family $\cal P$ of complete sets of orthogonal projectors, and
 the simplex $\triangle$ representing all possible spectra,
\begin{equation}
S={\cal P} \times \triangle .
\label{set}
\end{equation}
On the other hand, any element of ${\cal P}$ can be represented
by a unitary transformation, and any element of $\triangle$,
as a diagonal matrix $D$ with the matrix elements
$\Lambda_{ij}=\delta_{ij} \Lambda_i$, such that
they fulfill $\sum_{i=1}^N \Lambda_i=1$. Such representation
 corresponds to the form:
\begin{equation}
 \varrho= U D U^{\dagger}.
\label {mac}
\end{equation}
Thus, a uniform distribution on the set  of all density matrices represented
by (\ref{set}) is constructed naturally  by postulating an uniform
distribution on unitary transformations $U(N)$ (the Haar measure),
and an uniform distribution on diagonal matrices $D$.

We have calculated numerically the volume of the set of matrices 
with positive partial transpose, and we have estimated in such a way the
volume of the separable states. An algorithm to generate random  $U(N)$ 
matrices was recently given in Refs. \cite{Karol1,Karol2}.
The random  diagonal matrix $D$ fulfills
that $\sum_{i=1}^N \Lambda_i=1$,
 so the vector $\tilde\Lambda=(\Lambda_1, \ldots,\Lambda_N)$ is localized on the
$N-1$ dimensional simplex $\triangle$ (see section II).
Physically speaking,
no component of this vector is  distinguished in any sense.
Random vectors $\tilde\Lambda$ are thus generated {\sl uniformly}
on this subspace according to the simple method described in the Appendix A.

The numeric algorithm is then straightforward:  firstly we generate random
 density matrices of any size $N=N_1\times N_2$, secondly we construct their
 partial transpositions and finally we diagonalize them and we check
whether their eigenvalues $\lambda_i; i=1,\dots,N$ are all positive.
This procedure has been repeated several millions times in order 
to obtain the accuracy of the order of $1/1000$.

We should recall that in previous sections we have obtained rigorous
analytical lower and upper bounds of $\mu ( S_{sep})$, i.e. we have 
proven that $0<\mu ( S_{sep})<1$. The lower bounds follow from the fact that the states
 sufficiently close to the totally mixed state
are separable for all $N$. We have also shown that states which have sufficiently
 large participation ratio (i.e. which are sufficiently 
mixed) have a positive partial transpose, and are thus separable for $N=4,6$.
 On the other hand, the upper bounds come from the fact that matrices with large
 eigenvalues corresponding to an entangled eigenvector, are necessarily inseparable.
 Let us denote the measure $\mu ( S_{sep})$ in
 the $N_1 \times N_2$ case by $P_{N_1 \times N_2}$.
Our analytical bounds for the 
cases $2\times 2$ and $2\times 3$ are summarize below:
\begin{eqnarray}
0.302 & <&P_{2\times2}   < 0.863\, ,\nonumber \\
0.056 &<& P_{2\times3}.
\end{eqnarray}
Our numerical results agree  with 
these bounds but to  our surprise the probability 
that a mixed state $\varrho \in  {\cal H}_2 \times {\cal H}_2$ is
separable exceeds $50\%$. The results are  
\begin{equation}
P_{2\times 2}\approx 0.632\pm
0.002 ~ ~ {\rm and} ~ ~ P_{2\times 3}
\approx 0.384 \pm 0.002.
\label{numres}
\end{equation}
  For higher dimensions our results are summarized in Figure 1. This figure displays the 
probability $P_N$ that the partially transposed
matrix $\varrho^{T_2}$ is positive as a function of $N=N_1 \times N_2$.
 For $N=4$ and $N=6$ this is just the required probability of
encountering a separable state, while for $N>6$ it gives an upper bound for this quantity.

Due to symmetry of the problem $P_{N_1\times N_2}$ must be equal to
$P_{N_2\times N_1}$. Numerical results strongly suggest that
this quantity depends only on the product $N=N_1\times N_2$,
eg. $P_{2\times 6}=P_{3\times 4}$.
Moreover, the dependence can be well reproduced
by an exponential decay. The best fit gives
$P_N\sim 1.8 e^{-0.26 N}$.
We conjecture, therefore, that the measure of separable states
decreases exponentially with the size of the system in consideration.

\subsection{Purity versus separability}

Here we would like to illustrate  the physical connection
between the participation ratio and entanglement, which was already
 discussed in Section III.C.  Recalling that the participation ratio:
 $R(\varrho)=1/{\rm {\rm Tr}}(\varrho^2)$,  gives a characterization
of the degree of mixture and can be interpreted as the effective number
of states on the mixture. We have demonstrated 
 that if the state $\varrho$ has a sufficiently
 large participation ratio, or equivalently a 
sufficiently low von Neuman--Renyi 
entropy $H_2(\varrho)=\ln R(\varrho)$, then it partial tranposed is always positive.
 This holds for any arbitrary $N$, so in particular it means it is separable for $N=4,6$.
 A more precise estimate can be performed numerically.

For example consider the $N=4$--dimensional Hilbert space. A manifold
of the constant participation ratio $R$ is given by the ellipsoid in the space of eigenvalues:
$$\Lambda_1^2+\Lambda_2^2+\Lambda_3^2+(1-\Lambda_1-\Lambda_2-\Lambda_3)^2=1/R.$$
The probability distribution $P(R)$ obtained numerically using the natural
uniform measure on the 3-dimensional simplex is
plotted in Fig 2(a). 
It corresponds to the relative volume  of the cross-section
of a 3D hypersphere of radius $R^{-1/2}$ centered at $(0,0,0,0)$
with the simplex defined by a condition,
$\Lambda_1+\Lambda_2+\Lambda_3+\Lambda_4=1$.
 For $N=4$ and $R>3$ we obtain
$P(R)=6\pi R^{-2}\sqrt{1/R-1/4}$.

We have generated randomly a million points in the 3D simplex, computed
the corresponding participation ratio, rotated the corresponding
state by a random unitary matrix $U$ and checked, whether the
generated state is separable. This procedure allows us to investigate
the dependence of the probability of the separable states on the participation
ratio.  Our numerical results are summarized in Fig 2 (b), and again,
are fully compatible with the theorems and lemmas obtained in 
the previous sections.

Similar results and compatibility
are obtained for the case $N=6$. In this case we deal with a 5D simplex.
 Numerical data, displayed in Fig.3,  support the general fact that the
quantum states with $R(\varrho)\ge N-1$ have positive partial transpose (i.e.
are separable for $N=4,6$)  in the $N$-dimensional Hilbert space.
It is interesting to note that the relative
amount of separable states unambiguously increases as the participation ratio
grows. Moreover, the mean 'degree of entanglement' $\langle t \rangle$,
defined in the Appendix B, decreases monotonically with $R$. 
 This illustrates  that for composite systems
there exist a {\it dualism}
between the two quantities: purity and separability of the state.
The purest a quantum state is, the smallest its probability of being separable.
This conclusion is supported by Fig. 4  where
we plot the dependence of the probability of
finding a separable state $P_{sep}$ on the values of several
quantum Renyi entropies $H_q(\varrho)$. For $q \ne 1$ the Renyi entropy 
is defined as: $H_q(\varrho):=(\ln [$Tr$\varrho^q])/(1-q)$, while in the limit
$q\to 1$ it tends to the standard von Neumann entropy 
$H_1(\varrho)=-$Tr$\varrho \ln\varrho$.
We have not been able to generalize
 the rigorous the results for $q=2$, for which large entropy 
implies necessarily positivity of the partial transpose.
 Nevertheless, the numerical results suggest that similar result
 holds for arbitrary values of the Renyi parameter $q$.
 All states with sufficiently
 large $H_q(\varrho)$ are separable as shown in figure 4a.
 Can this fact be used 
to obtain a better lower bound for $P_{N_1\times N_2}$?
 Fig. 4b suggests that it is not the case: the cumulative probability
$P$ of $H_q(\varrho)$
attains more or less the same value for all $q$ at the point at which  
$P_{sep}$ becomes $\simeq 1$.

\section{Conclusions}

Summarizing, we have developed a measure theoretical approach to the 
separability--inseparability problem. To this aim, we have proposed a
natural measure in the space of density matrices $\varrho$ on the 
$N$--dimensional space. We have proven that,
under this measure, the set of separable states has
a nonzero volume although this volume is not maximal in the set of all states. 
Analytical lower and upper bounds of this volume have been found  
for $N=2 \times 2$, and $N=2 \times 3 $ cases.
We have also provided qualitative evidence that
for $N\ge 8$ the peculiar set of inseparable states with positive partial
transposition has, under this measure, a nonzero volume.

We have used Monte Carlo simulations to estimate with much higher 
precision the volume of separable states. Our numerical simulations give strong 
evidence that this volume decreases exponentially with the 
dimension of the composite quantum system. 
Finally, we have also discussed the dualism between
purity and separability, and have shown that while
entanglement is typical of pure states,
separability is rather connected with quantum mixtures. 

Several questions concerning this subject remain still as open problems 
and so far, we have not been able to prove them rigorously.
Particularly challenging are the two following related questions:
\begin{itemize}
\item Does the volume of the set of separable states goes
 really to zero as the dimension of the composite system $N$ grows, and how fast?

\item{} Has the set of separable states really a volume
strictly smaller than the volume of the set of states with
a positive partial transpose?
\end{itemize}

\vskip 0.2cm
{\bf Acknowledgments}

We thank Guifr\'e Vidal for very illuminating ideas concerning Section V.
It is also a pleasure to thank D. DiVincenzo,  M. Horodecki,  S. Popescu
and W.H. {\.Z}urek for fruitful discussions.
One of us (K.{\.Z}) would like to thank for the hospitality to
Isaac Newton Institute for Mathematical Science in Cambridge where a
part of this work has been  done. P.H. is grateful to Fundacja Nauki
Polskiej for a financial support and 
A.S. kindly  acknowldges a support from
CEE and DGICYT (Spain) under a contract number PB95-0778-C02-02.

\appendix

\section{Generation of  uniform distribution on the simplex}

Our aim is to construct the uniform distribution of the points {\sl on}
the manifold given by
 $\sum_{i=1}^N \Lambda_i=1$, where each component $\Lambda_i$ is non negative.
 Let $\xi_i; i=1,\dots,N-1$ be independent random
numbers generated uniformly in the interval $(0,1)$.
We start with a uniform
 distributions {\sl inside} the $N-1$
dimensional simplex $\Delta_{n-1}$ defined by $\sum_{i=1}^{N-1} \Lambda_i < 1$.
Its  volume is proportional to the product
$\prod_{k=1}^{N-1} x_k^{N-1-k} dx_k=
\prod_{k=1}^{N-1} d[x_k^{N-k}]$, what enables us to
find the required densities for each component.
Since the vertex of the simplex $\Delta_{N-1}$ is situated at
$\{0,\dots,0\}$,
the largest weight corresponds to the small values of $x_1$. Therefore:
$$ \Lambda_1=1-\xi_1^{1\over N-1}, $$
$$ \Lambda_2=[1-\xi_2^{1\over N-2}](1-\Lambda_1), $$
$$ \Lambda_k=[1-\xi_k^{1\over N-k}](1-\sum_{i=1}^{k-1} \Lambda_i), $$
$$\cdots$$
$$ \Lambda_{N-1}=[1-\xi_{N-1}](1-\sum_{i=1}^{N-2} \Lambda_i). $$
Eventually the  last component $\Lambda_N$ is already determined as:
$$\Lambda_N=1- \sum_{i=1}^{N-1} \Lambda_i.$$
 The vector $\tilde\Lambda=\{\Lambda_1,\dots,\Lambda_N\}$ constructed in this way
is distributed uniformly in the requested subspace.
An alternative procedure, albeit more time consuming, is to take 
any vector of an $N\times N$ auxiliary random unitary matrix $V$
and obtain the random vector as $\Lambda_k=|V_{kj}|^2$ 
with arbitrary  $j$.

\section{The averaged "degree of entanglement"}

The problem of defining a quantity capable to measure a 
"degree of entanglement" is a subject of several recent 
studies \cite{VPRK97,PR97,VPJK97,VP98}. 
Let us define for a given density matrix $\varrho$ the quantity
$$ t:=\sum_{i=1}^N |\lambda'_i|-1, $$
where
 $\lambda'_i,i=1,\dots,N$ denote the eigenvalues of the partially
transposed  matrix $\varrho^{T_2}$.
 For any separable matrix all eigenvalues are positive,
 its trace is equal to unity and $t$ equals to zero. 
On the other hand, for the maximally entangled states belonging to 
$2\times 2$ system the spectrum of eigenvalues  $\lambda'$
consists of $\{-1/2,1/2,1/2,1/2\}$, so that
$t=1$. Moreover, for the often studied $2\times2$ Werner states \cite{Werner}
depending on the parameter $x$, the quantity $t$ vanishes for $x<2/3$
(separable states) and equals to $t=(3x-2)/(4-3x)$ for entangled
states ($2/3 \le x\le 1$).

We could not resist the temptation to investigate the mean value of
$t$ averaged over random density matrices generated as described above.
For the $2\times 2$ problem the mean value $\langle t\rangle$ equals to
$0.057$ and increases to $0.076$ for the $2\times 3$ problem.
For large systems this quantity seems to saturate at $t\sim 0.10$,
as the ratio of the matrices with positive values of $t$
(some eigenvalues of $\varrho^{T_2}$ are negative) tends to unity.
Moreover, as shown in Fig.2b and Fig 3b, the average degree of
entanglement $\langle t\rangle$ 
decreases monotonically with the participation ratio $R$,
what provides a quantitative characterization of the relation 
between entanglement and purity of mixed quantum states.

\begin{figure}
\caption{Probability of finding a state with positive partial transpose 
as a function of
the dimension of the problem $N$; For $N>6$ it gives an upper
bound only of the relative volume of the separable states.
Different symbols distinguish different sizes of one subsystem 
($k=$ $2(\diamond)$, $3 (\triangle)$ and $4 (\times)$),
while the solid line represents the exponential fit.}
\label{fig1}
\end{figure}

\begin{figure}
\caption{Purity and separability in $N$=4--dimensional Hilbert space: a) probability
of finding quantum state with a participation ratio $R$; b)
probability of finding a separable state $P_{2\times 2}$ as a function of the
participation ratio $R$ (crosses). All states beyond the dashed vertical
line placed at $R=N-1$  are separable.
Circles show the mean entanglement $\langle t \rangle$, as defined in
Appendix B. }
\label{fig2}
\end{figure}

\begin{figure}
\caption{Same as in Fig. 2 for  $N=6$.}
\label{fig3}
\end{figure}

\begin{figure}
\caption{ (a) Probability of separable states $P_{2\times 2}$ for
$N=4$  vs von Neumann-Renyi entropies $H_q \in [0,\ln(4)]$ for 
 $q=1 (\star )$, $ 2 (\circ )$,
$3 (\triangle )$, and $ 10 (\diamond )$.
 (b)  Integrated  distribution function $D(H_q)$. 
Vertical line drawn at $D\sim 0.7$ corresponds to these values
of $H_q$ for which $P_{sep}=1$.  } 
\label{fig4}
\end{figure}

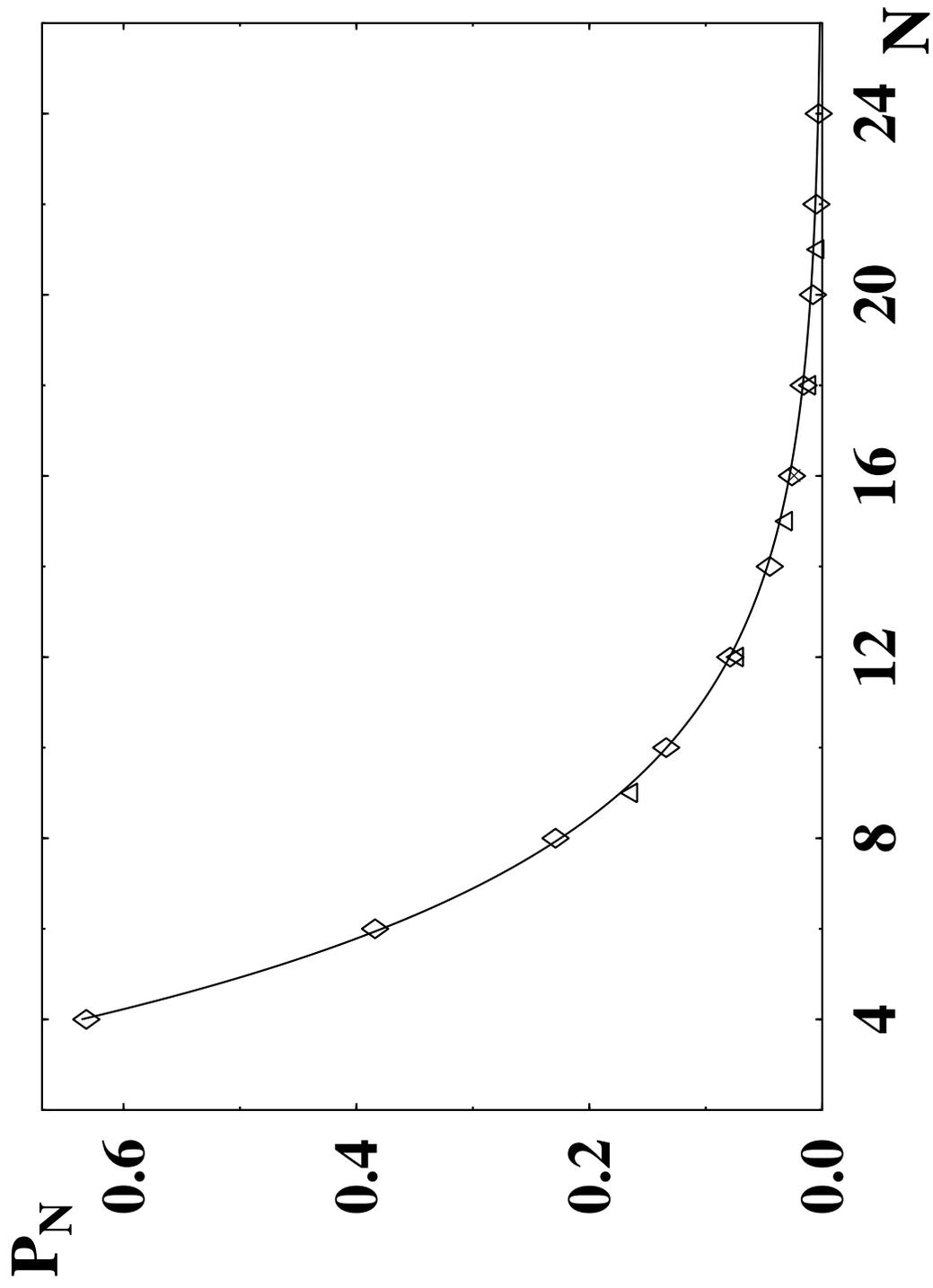
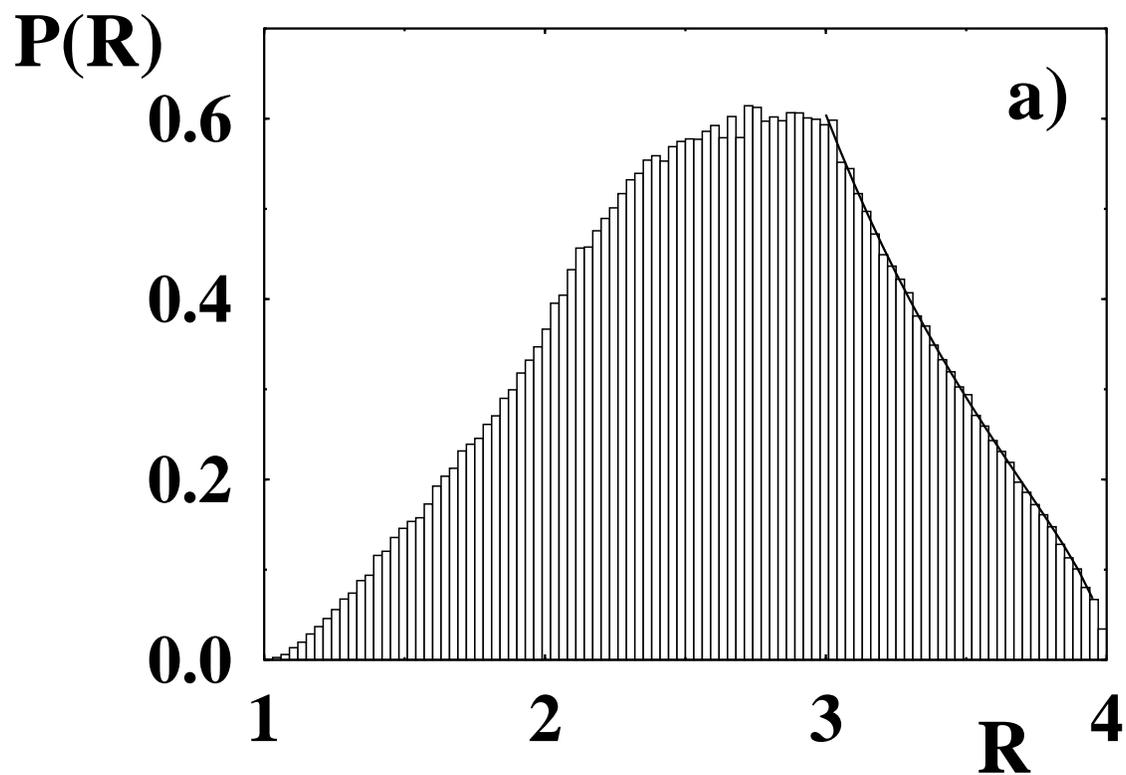
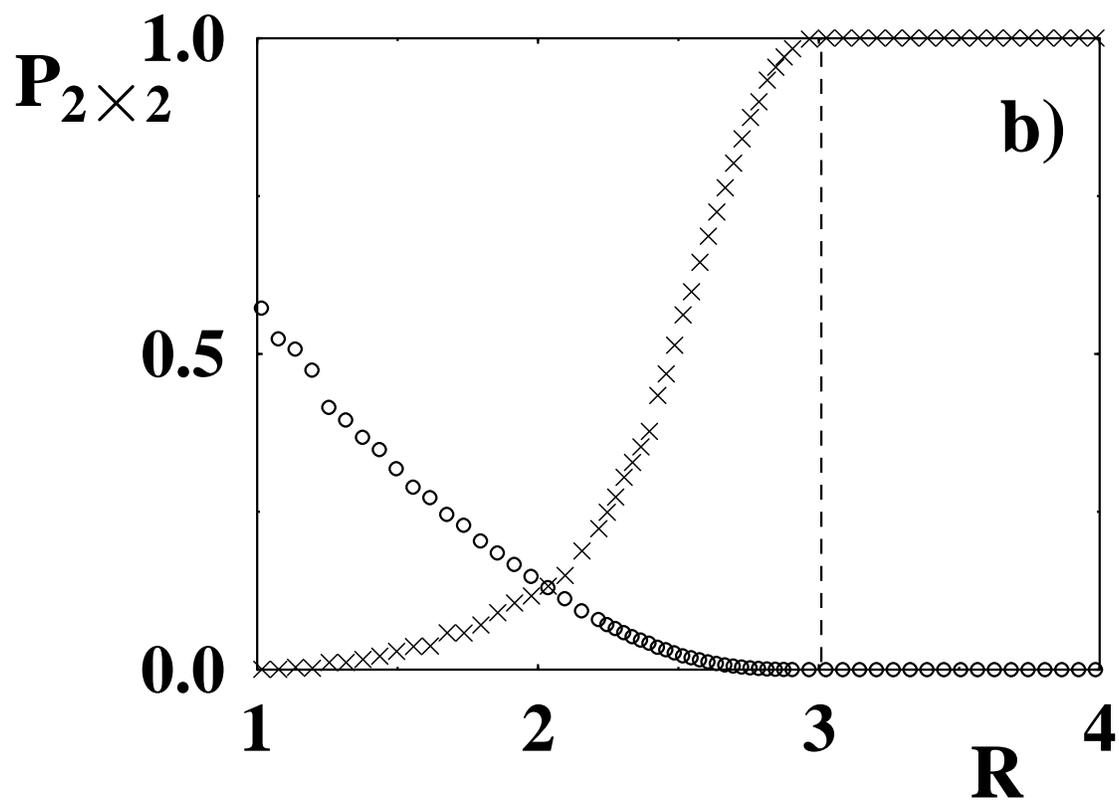
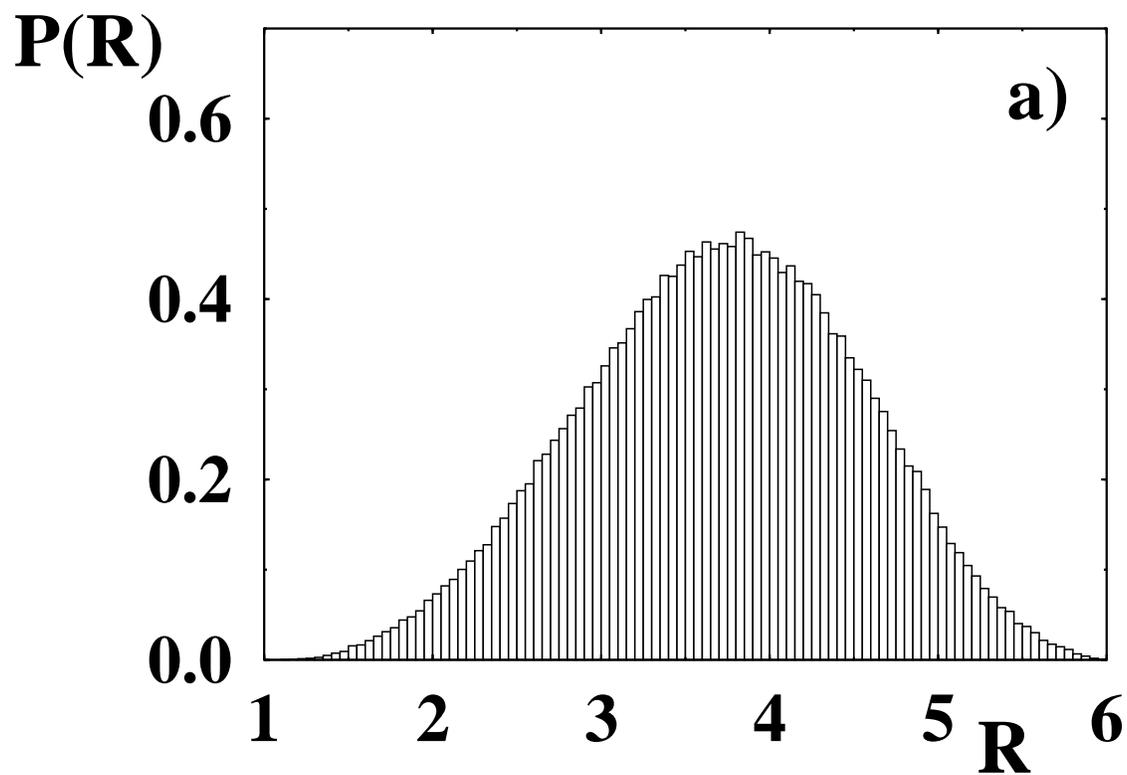
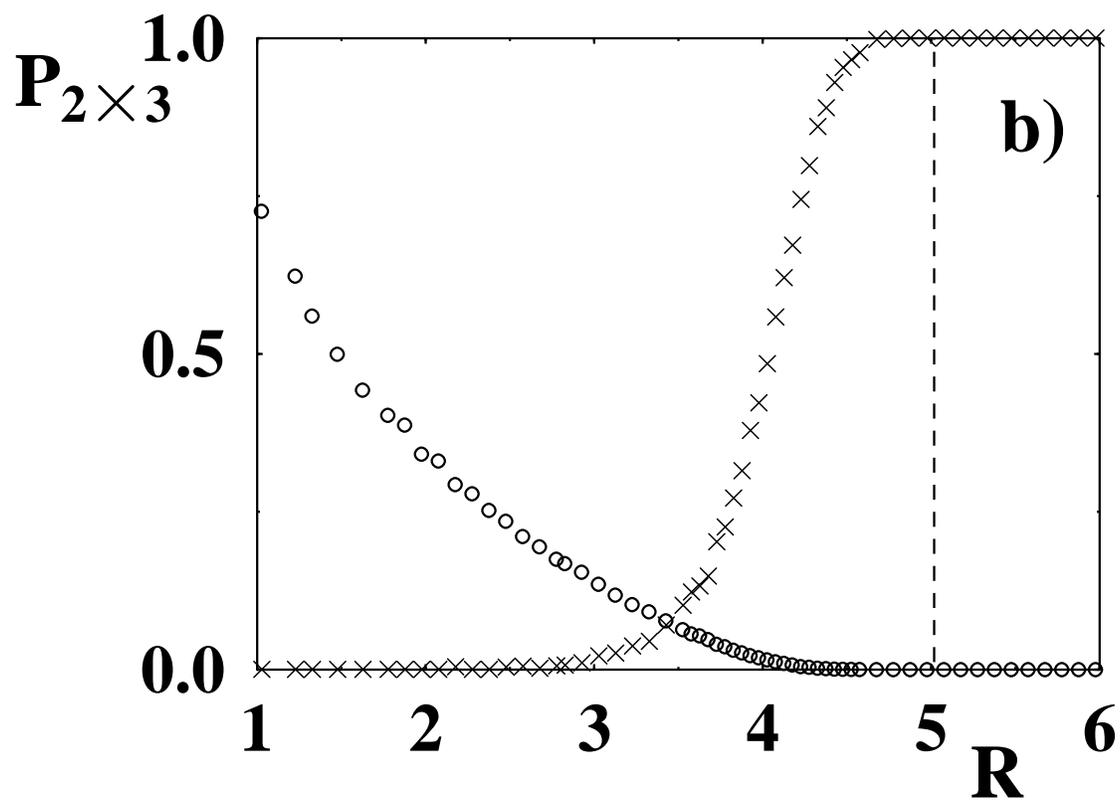
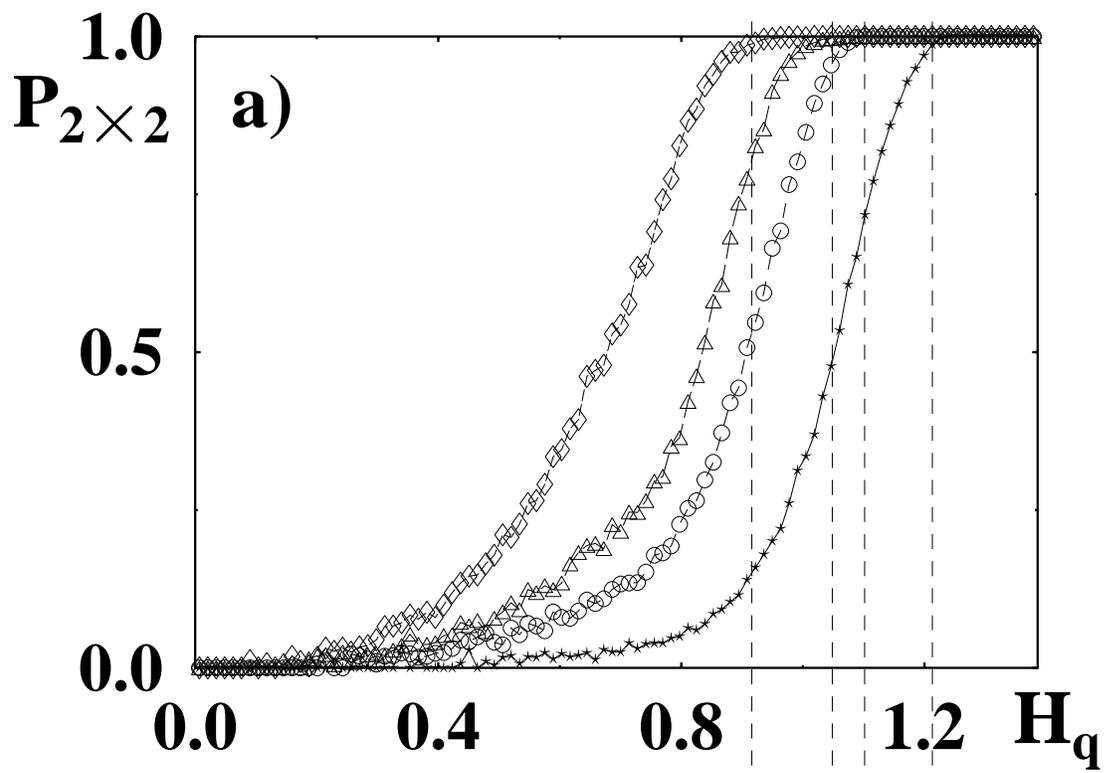
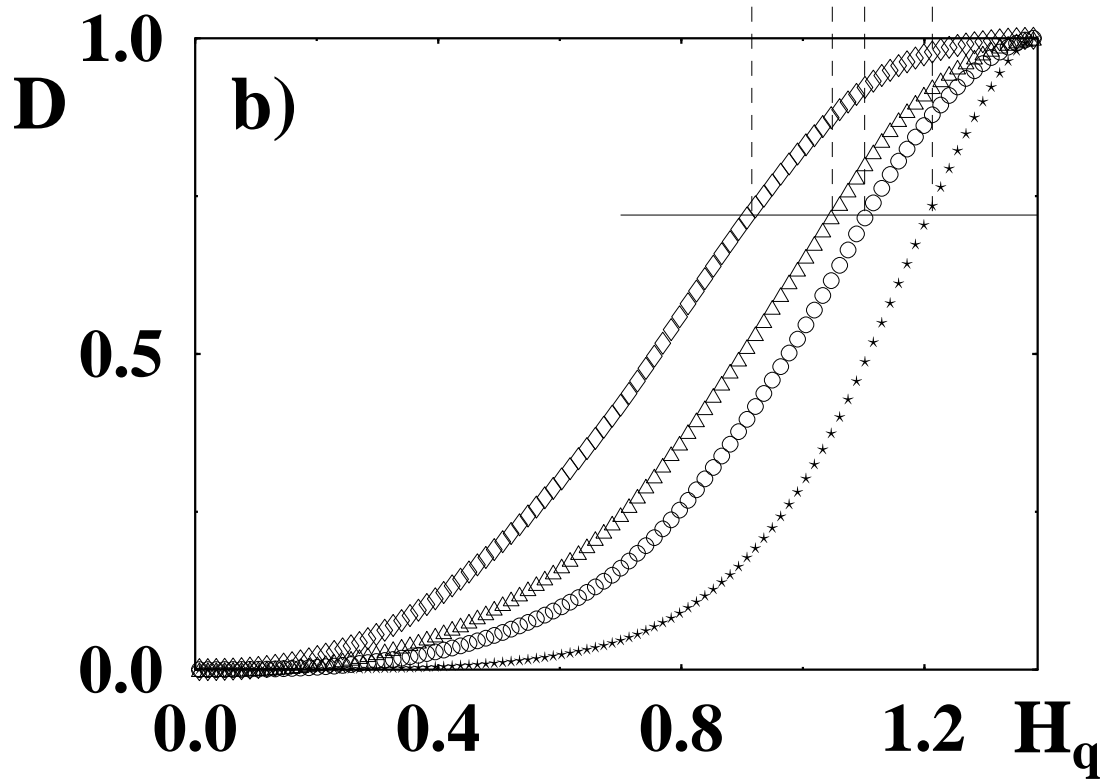
\end{document}